\documentclass[a4paper,11pt]{article}
\pdfoutput=1
\usepackage{jheppub}
\newcommand{\Li}[1]{\mathop{\mathrm{Li}}\nolimits_{#1}}
\newcommand{\F}[4]{\,_{#1}F_{#2}\left(\left.\begin{array}{c}#3\end{array}\right|#4\right)}
\DeclareMathOperator{\Tr}{Tr}
\preprint{\begin{flushright}MITP/18-037\end{flushright}}
\title{Four-loop cusp anomalous dimension in QED}
\author{Andrey Grozin}
\affiliation{PRISMA Cluster of Excellence,
Johannes Gutenberg University,\\
Staudingerweg 9, 55128 Mainz, Germany;\\
Budker Institute of Nuclear Physics SB RAS,
Lavrentyev st.~11, Novosibirsk 630090, Russia;\\
Novosibirsk State University,
Pirogova st.~2, Novosibirsk 630090, Russia}
\emailAdd{A.G.Grozin@inp.nsk.su}

\abstract{The 4-loop $C_F^3 T_F n_l$ and 5-loop $C_F^4 T_F n_l$ terms
in the HQET field anomalous dimension $\gamma_h$ are calculated analytically
(the 4-loop one agrees with the recent numerical result~\cite{Marquard:2018rwx}).
The 4-loop $C_F^3 T_F n_l$ and 5-loop $C_F^4 T_F n_l$ terms
in the cusp anomalous dimension $\Gamma(\varphi)$ are calculated analytically,
exactly in $\varphi$ (the $\varphi\to\infty$ asymptotics of the 4-loop one
agrees with the recent numerical result~\cite{Moch:2017uml}).
Combining these results with the recent 4-loop $d_{FF} n_l$ contributions to $\gamma_h$
and to the small-$\varphi$ expansion of $\Gamma(\varphi)$ up to $\varphi^4$~\cite{Grozin:2017css}
(recently extended to $\varphi^6$~\cite{Bruser:2018aud})
we now have the complete analytical 4-loop result
for the Bloch--Nordsieck field anomalous dimension in QED,
and the small-$\varphi$ expansion of the 4-loop QED cusp anomalous dimension up to $\varphi^6$.}
\keywords{QCD, QED, Effective field theories}

\begin{document}
\maketitle

\section{Introduction}
\label{S:Intro}

QCD problems with a single heavy quark $Q$ having momentum $P = M_Q v + p$
(where $M_Q$ is the on-shell mass and $v$ is some vector with $v^2 = 1$)
can be described by heavy quark effective theory (HQET)
if characteristic heavy-quark residual momentum $p$,
as well as characteristic gluon and light-quark momenta $k_i$,
are $\ll M_Q$ (see, e.g., \cite{Manohar:2000dt,Grozin:2004yc,Grozin:2013bva}).
The heavy quark is described by the field
\begin{equation}
h_{v0} = Z_h^{1/2}(\alpha_s(\mu),a(\mu)) h_v(\mu)\,,
\label{Intro:h}
\end{equation}
where we use the $\overline{\text{MS}}$ scheme,
and $Z_h$ is a minimal renormalization constant.
We use the covariant gauge:
$-(\partial_\mu A_0^{\mu a})^2/(2 a_0)$ is added to the Lagrangian,
the gauge-fixing parameter is renormalized by the same minimal constant
as the gluon field: $a_0 = Z_A(\alpha_s(\mu),a(\mu)) a(\mu)$.
The HQET heavy-quark field anomalous dimension
is defined as $\gamma_h = d\log Z_h/d\log\mu$.
The $h_{v0}$ coordinate-space propagator in the $v$ rest frame has the form
\begin{equation}
S_h(x) = \delta^{(d-1)}(\vec{x}) \theta(x^0) W(x^0)\,,
\label{Intro:Sh}
\end{equation}
where $W(t)$ is the straight Wilson line along $v$ of length $t$.
The heavy-quark field is QCD and HQET are related
by the matching coefficient $z$~\cite{Grozin:2010wa}:
\begin{equation}
Q_0 = z_0^{1/2} h_{v0} + \mathcal{O}(1/M_Q)\,,\quad
Q(\mu) = z^{1/2}(\mu) h_v(\mu) + \mathcal{O}(1/M_Q)\,.
\label{Intro:z}
\end{equation}

The HQET field anomalous dimension $\gamma_h$
is known up to three loops~\cite{Melnikov:2000zc,Chetyrkin:2003vi}.
In the first of these papers, it was obtained as a by-product of the three-loop calculation
of the heavy-quark field renormalization constant in the on-shell scheme $Z_Q^{\text{os}}$,
from the requirement that the renormalized matching coefficient $z(\mu)$~(\ref{Intro:z})
must be finite;
in the second paper, it was confirmed by a direct HQET calculation.
Several color structures of the 4-loop result are also known:
$C_F (T_F n_l)^3$~\cite{Broadhurst:1994se} ($n_l$ is the number of light flavors),
$C_F^2 (T_F n_l)^2$~\cite{Grozin:2015kna,Grozin:2016ydd},
$C_F C_A (T_F n_l)^2$~\cite{Marquard:2018rwx}
(from the analytical $Z_Q^{\text{os}}$ result~\cite{Lee:2013sx} using the finiteness of $z(\mu)$),
$d_{FF} n_l$~\cite{Grozin:2017css}.
Here $C_R$ ($R=F$, $A$) are the standard quadratic Casimirs: $t_R^a t_R^a = C_R \mathbf{1}_R$
($t_R^a$ are the generators in the representation $R$,
$\mathbf{1}_R$ is the corresponding unit matrix);
$\Tr t_F^a t_F^b = T_F \delta^{ab}$;
$d_{FF} = d_F^{abcd} d_F^{abcd} / N_c$,
where $d_R^{abcd} = \Tr t_R^{(a} t_R^b t_R^c t_R^{d)}$ (the brackets mean symmetrization),
and $N_c = \Tr\mathbf{1}_F$.
The remaining terms are known numerically~\cite{Marquard:2018rwx},
from the numerical 4-loop $Z_Q^{\text{os}}$ using the finiteness of $z(\mu)$~(\ref{Intro:z}).
Here I calculate the $C_F^{L-1} T_F n_l \alpha_s^L$ terms up to $L=5$
analytically (sect.~\ref{S:gammah});
the $L=4$ term agrees with the numerical result~\cite{Marquard:2018rwx}.

If the heavy-quark velocity is substantially changed
(e.\,g., a weak decay into another heavy quark),
we have HQET with 2 unrelated fields $h_v$, $h_{v'}$.
At the effective-theory level this is described by the current
\begin{equation}
J_0 = h^+_{v' 0} h_{v 0} = Z_J(\alpha_s(\mu)) J(\mu)\,.
\label{Intro:J}
\end{equation}
The minimal renormalization constant $Z_J$ is gauge invariant (unlike $Z_h$)
because the current $J_0$ is color singlet.
The anomalous dimension of this current,
also known as the cusp anomalous dimension,
is defined as $\Gamma(\varphi) = d\log Z_J/d\log\mu$,
where $\cosh\varphi = v\cdot v'$.

The QCD cusp anomalous dimension $\Gamma(\varphi)$
is known up to three loops~\cite{Grozin:2014hna,Grozin:2015kna}.
At $\varphi\ll1$ it is a regular series in $\varphi^2$.
At $\varphi\gg1$ it is $\Gamma_l \varphi + \mathcal{O}(\varphi^0)$~\cite{Korchemsky:1987wg},
where $\Gamma_l$ is the light-like cusp anomalous dimension.
Several color structures of the 4-loop $\Gamma(\varphi)$ are also known:
$C_F (T_F n_l)^3$~\cite{Beneke:1995pq},
$C_F^2 (T_F n_l)^2$~\cite{Grozin:2015kna,Grozin:2016ydd}.
The $d_{FF} n_l$ term is known at $\varphi\ll1$ up to $\varphi^4$~\cite{Grozin:2017css}%
\footnote{Such expansion was first used in~\cite{Bagan:1993js} at 2 loops.}.
For the $\varphi\gg1$ asymptotics (i.\,e. $\Gamma_l$), both $n_l^2$ terms are known
from combining the $C_F^2 (T_F n_l)^2$ result~\cite{Grozin:2015kna,Grozin:2016ydd}
and the large-$N_c$ $N_c^2 n_l^2$ result~\cite{Henn:2016men}.
Large-$N_c$ results for $\Gamma_l$ at $n_l^1$~\cite{Henn:2016men,Moch:2017uml}
and $n_l^0$~\cite{Lee:2016ixa,Moch:2017uml} are also known analytically.
Contributions of individual color structures of $\Gamma_l$ at $n_l^{1,0}$
are only known numerically~\cite{Moch:2017uml}.
Here I calculate the $C_F^{L-1} T_F n_l \alpha_s^L$ terms up to $L=5$ in $\Gamma(\varphi)$
analytically, as exact functions of $\varphi$ (sect.~\ref{S:Gamma}).
In particular, I find their $\varphi\gg1$ asymptotics;
the analytical $L=4$ result agrees with the numerical one~\cite{Moch:2017uml}.

In QED without light lepton flavors ($n_l=0$), as explained below,
both $\gamma_h$ and $\Gamma(\varphi)$ are exactly given by the one-loop formulas.
When $n_l\ne0$, higher corrections appear.
Combining the 4-loop $\gamma_h$ results for
$C_F (T_F n_l)^3$~\cite{Broadhurst:1994se},
$C_F^2 (T_F n_l)^2$~\cite{Grozin:2015kna,Grozin:2016ydd},
$C_F^3 T_F n_l$ (sect.~\ref{S:gammah})
and $d_{FF} n_l$~\cite{Grozin:2017css} structures,
I obtain the complete analytical 4-loop result
for the Bloch--Nordsieck field anomalous dimension $\gamma_h$ in QED (sect.~\ref{S:QED}).
Combining the 4-loop $\Gamma(\varphi)$ full results for
$C_F (T_F n_l)^3$~\cite{Beneke:1995pq},
$C_F^2 (T_F n_l)^2$~\cite{Grozin:2015kna,Grozin:2016ydd},
$C_F^3 T_F n_l$ (sect.~\ref{S:Gamma}) structures
with the $d_{FF} n_l$ term~\cite{Grozin:2017css} (expansion up to $\varphi^4$),
I obtain the expansion of the 4-loop QED $\Gamma(\varphi)$
up to $\varphi^4$ (sect.~\ref{S:QED}).

\section{HQET field anomalous dimension: the $C_F^{L-1} T_F n_l \alpha_s^L$ terms}
\label{S:gammah}

\begin{figure}[ht]
\begin{center}
\begin{picture}(94,20)
\put(21,9.75){\makebox(0,0){\includegraphics{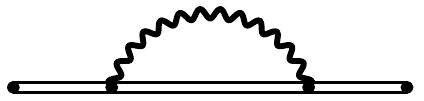}}}
\put(73,12.5){\makebox(0,0){\includegraphics{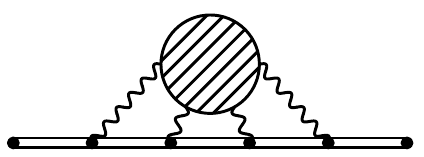}}}
\put(21,0){\makebox(0,0)[b]{a}}
\put(73,0){\makebox(0,0)[b]{b}}
\end{picture}
\end{center}
\caption{Webs: (a) 2-leg (the thick line is the full photon propagator);
(b) 4-leg (the blob is the sum of connected diagrams).}
\label{F:webs}
\end{figure}

This is a QED problem.
Due to exponentiation~\cite{Gatheral:1983cz,Frenkel:1984pz},
the coordinate-space propagator of the Bloch--Nordsieck field
(i.\,e.\ the straight Wilson line $W$) is
\begin{equation}
W = \exp \left( \sum w_i \right)\,,
\label{gammah:exp}
\end{equation}
where $w_i$ are single-web diagrams.
Due to $C$ parity conservation in QED, webs have even numbers of legs (fig.~\ref{F:webs}).
In QED with $n_l=0$ there is only 1 single-web diagram:
fig.~\ref{F:webs}a with the free photon propagators.
Therefore, $\log W$ is exactly 1-loop;
the $\beta$ function is 0, and hence $\gamma_h$ is also exactly 1-loop.
At $n_l>0$ corrections to the photon propagator in fig.~\ref{F:webs}a appear.
Webs with 4 legs (fig.~\ref{F:webs}b) first appear at 4 loops;
they have been calculated in~\cite{Grozin:2017css}.
All contributions to $\log W$~(\ref{gammah:exp})
are gauge invariant except the 1-loop one,
because proper vertex functions with any numbers of photon legs
are gauge invariant and transverse with respect to each photon leg
due to the QED Ward identities.

The full momentum-space photon propagator in the covariant gauge is
\begin{equation}
D^{\mu\nu}(k) = - \frac{i}{k^2}
\left(g^{\mu\nu} - \frac{k^\mu k^\nu}{k^2}\right)
\frac{1}{1 - \Pi(k^2)}
- i a_0 \frac{k_\mu k_\nu}{(k^2)^2}\,,
\label{gammah:D}
\end{equation}
where $\Pi(k^2)$ is the photon self-energy:
\begin{equation}
\Pi = \sum_{L=1}^\infty \Pi_L A_0^L (-k^2)^{-L\varepsilon}\,,\quad
A_0 = \frac{e_0^2}{(4\pi)^{d/2}} e^{-\gamma\varepsilon}
\label{gammah:Pi}
\end{equation}
($e_0^2$ has dimensionality $m^{2\varepsilon}$,
so that the power of $-k^2$ is obvious;
$\gamma$ is the Euler constant).
Only the 0-loop term in~(\ref{gammah:D}) is gauge dependent.
Writing $\Pi_L$ as $\tilde{\Pi}_L n_l + (n_l^{>1} \text{ terms})$,
we obtain in the Landau gauge $a_0=0$
\begin{align}
&D^{\mu\nu}(k) = \tilde{D}_0^{\mu\nu}(k)
+ n_l \sum_{L=1}^\infty \tilde{\Pi}_L \tilde{D}_L^{\mu\nu}(k) A_0^L
+ (n_l^{>1} \text{ terms})\,,
\nonumber\\
&\tilde{D}_L^{\mu\nu}(k) = \frac{i}{(-k^2)^{1+L\varepsilon}}
\left(g^{\mu\nu} + \frac{k^\mu k^\nu}{-k^2}\right)\,.
\label{gammah:Dk}
\end{align}
The $\overline{\text{MS}}$ charge renormalization is
\begin{align}
&A_0 = \mu^{2\varepsilon} \frac{\alpha(\mu)}{4\pi} Z_\alpha(\alpha(\mu))\,,
\label{gammah:ren}\\
&\frac{d\log\alpha(\mu)}{d\log\mu} = - 2 \varepsilon - 2 \beta(\alpha(\mu))\,,\quad
\beta(\alpha) = \frac{1}{2} \frac{d\log Z_\alpha}{d\log\mu}
= \sum_{L=1}^\infty \beta_L \left(\frac{\alpha}{4\pi}\right)^L
\nonumber
\end{align}
(note that here we call the $L$-loop $\beta$ function coefficient $\beta_L$,
not $\beta_{L-1}$ as usually done; this makes subsequent formulas more logical).
In QED $\log\left(1 - \Pi(k^2)\right) = \log Z_\alpha + (\text{finite})$;
writing $\beta_L = \bar{\beta}_L n_l + (n_l^{>1} \text{ terms})$,
we see that $1/\varepsilon$ terms in $\tilde{\Pi}_L$ are related to $\bar{\beta}_L$:
\begin{equation}
\tilde{\Pi}_L = \frac{\bar{\beta}_L}{L} \frac{1}{\varepsilon} + \bar{\Pi}_L
+ \mathcal{O}(\varepsilon)\,.
\label{gammah:Pieps}
\end{equation}
Here the $\beta$ function coefficients are~\cite{Gorishnii:1990kd}
\begin{equation}
\bar{\beta}_1 = - \frac{4}{3}\,,\quad
\bar{\beta}_2 = - 4\,,\quad
\bar{\beta}_3 = 2\,,\quad
\bar{\beta}_4 = 46\,,
\label{gammah:Pi2}
\end{equation}
and~\cite{Ruijl:2017eht}
\begin{align}
&\bar{\Pi}_1 = - \frac{20}{9}\,,\quad
\bar{\Pi}_2 = 16 \zeta_3 - \frac{55}{3}\,,\quad
\bar{\Pi}_3 = - 2 \left(80 \zeta_5 - \frac{148}{3} \zeta_3 - \frac{143}{9}\right)\,,
\nonumber\\
&\Bar{\Pi}_4 = 2240 \zeta_7 - 1960 \zeta_5 -104 \zeta_3 + \frac{31}{3}\,.
\label{gammah:Pires}
\end{align}

The coordinate-space full photon propagator is the Fourier transform of~(\ref{gammah:Dk}):
\begin{align}
&D^{\mu\nu}(x) = \frac{1}{(4\pi)^{d/2}} \left[ \bar{D}_0^{\mu\nu}(x)
+ n_l \sum_{L=1}^\infty \tilde{\Pi}_L \bar{D}_L^{\mu\nu}(x) A_0^L \right]
+ (n_l^{>1} \text{ terms})\,,
\label{gammah:Dx}\\
&\bar{D}_L^{\mu\nu}(x) =
\frac{\Gamma\bigl(1-(L+1)\varepsilon\bigr)}{\Gamma\bigl(1+L\varepsilon\bigr)}
\left(\frac{4}{-x^2}\right)^{1-(L+1)\varepsilon}
\nonumber\\
&\qquad{}\times\left[- g^{\mu\nu} + \frac{g^{\mu\nu} + 2 \bigl(1 - (L+1) \varepsilon\bigr) x^\mu x^\nu / (-x^2)}{2 (1+L\varepsilon)}\right]\,.
\nonumber
\end{align}

The sum of single-web diagrams (fig.~\ref{F:webs}) in the Landau gauge,
analytically continued to Euclidean $t=-i\tau$, is
\begin{equation}
\log W = S_1 A
+ n_l \sum_{L=2}^\infty S_L \tilde{\Pi}_{L-1} A^L
+ (n_l^{>1} \text{ terms}) + (w_{>2 \text{ legs}} \text{ terms})\,,\quad
A = A_0 \left(\frac{\tau}{2}\right)^{2\varepsilon} e^{2\gamma\varepsilon}\,,
\label{gammah:wn}
\end{equation}
where the 1-loop HQET integral
\begin{equation}
S_L = \frac{3 - 2\varepsilon}{L\varepsilon (1 - 2 L\varepsilon) (1 + (L-1)\varepsilon)}
\frac{\Gamma(1 - L\varepsilon)}{\Gamma(1 + (L-1)\varepsilon)}
e^{- (2L-1) \gamma \varepsilon}
= \frac{3}{L\varepsilon} + 3 + \frac{1}{L} + \mathcal{O}(\varepsilon)
\label{gammah:S}
\end{equation}
can be calculated in coordinate space~(\ref{gammah:Dx}),
or as a Fourier transform of the momentum-space HQET propagator.
Now we re-express $\log W$ via the renormalized $\alpha(\mu)$ at $\mu_0 = 2 e^{-\gamma} / \tau$
(it is sufficient to do this in the 1-loop term) and obtain
\begin{align}
&\log W = S_1 \frac{\alpha}{4\pi} + n_l \sum_{L=2}^\infty
\left[S_L \left(\frac{\bar{\beta}_{L-1}}{L-1} \frac{1}{\varepsilon} + \bar{\Pi}_{L-1}\right) - S_1 \frac{\bar{\beta}_{L-1}}{L-1} \frac{1}{\varepsilon}\right]
\left(\frac{\alpha}{4\pi}\right)^L
\nonumber\\
&{} + (n_l^{>1} \text{ terms}) + (w_{>2 \text{ legs}} \text{ terms})
= \log Z_h + (\text{finite})\,.
\label{gammah:logW}
\end{align}
Extracting $\log Z_h$ and differentiating it in $\log\mu$, we obtain $\gamma_h$.
Restoring the color factors and adding the gauge dependent term%
\footnote{In the arbitrary covariant gauge the extra term to be added to $w_1$~(\ref{gammah:wn}) is
$\Gamma(-\varepsilon) e^{-\gamma\varepsilon} a_0 A = \Gamma(-\varepsilon) e^{-\gamma\varepsilon} a(\mu_0) \alpha(\mu_0)/(4\pi)$,
because in QED $Z_\alpha = Z_A^{-1}$.
Hence the extra term to be added to $\log Z_h$ is purely 1-loop:
$- (a/\varepsilon) \alpha/(4\pi)$.
In QED $d\log\left(a(\mu) \alpha(\mu)\right)/d\log\mu = -2\varepsilon$ exactly,
and hence the extra term in $\gamma_h$~(\ref{gammah:gamma}) is also purely 1-loop:
$2 a \alpha/(4\pi)$.},
we obtain
\begin{align}
\gamma_h ={}& \frac{\alpha_s}{4\pi} \biggl[ 2 (a-3) C_F + T_F n_l
\sum_{L=1}^\infty \bigl(- 6 \bar{\Pi}_L + 2 \bar{\beta}_L\bigr) \left(C_F \frac{\alpha_s}{4\pi}\right)^L \biggr]
+ (\text{other color structures})
\nonumber\displaybreak\\
{}={}& 2 (a-3) C_F \frac{\alpha_s}{4\pi} + T_F n_l C_F \left(\frac{\alpha_s}{4\pi}\right)^2 \biggl[
\frac{32}{3}
- 6 \left(16 \zeta_3 - 17\right) C_F \frac{\alpha_s}{4\pi}
\nonumber\\
&{} + \frac{16}{3} \left(180 \zeta_5 - 111 \zeta_3 - 35\right) \left(C_F \frac{\alpha_s}{4\pi}\right)^2\nonumber\\
&{} - 6 (2240 \zeta_7 - 1960 \zeta_5 - 104 \zeta_3 - 5) \left(C_F \frac{\alpha_s}{4\pi}\right)^3
+ \mathcal{O}(\alpha_s^4) \biggr]
\nonumber\\
&{} + (\text{other color structures})\,.
\label{gammah:gamma}
\end{align}
We have reproduced the $C_F^2 T_F n_l$ term in the 3-loop anomalous dimension~\cite{Melnikov:2000zc,Chetyrkin:2003vi}
by a simpler method.
The coefficient of $C_F^3 T_F n_l (\alpha_s/\pi)^4$ in $\frac{1}{2} \gamma_h$ is
\begin{equation*}
\frac{180 \zeta_5 - 111 \zeta_3 - 35}{96} \approx 0.189778\,,
\end{equation*}
in perfect agreement with the numerical result $0.1894\pm0.0030$ (Table~III in~\cite{Marquard:2018rwx}).

\section{QCD cusp anomalous dimension: the $C_F^{L-1} T_F n_l \alpha_s^L$ terms}
\label{S:Gamma}

\begin{figure}[ht]
\begin{center}
\begin{picture}(52,26)
\put(26,12){\makebox(0,0){\includegraphics{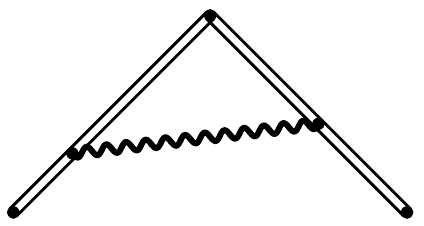}}}
\put(26,24){\makebox(0,0){$0$}}
\put(1.5,2){\makebox(0,0){$-vt$}}
\put(6.5,9){\makebox(0,0){$-vt_1$}}
\put(49,2){\makebox(0,0){$v't$}}
\put(41,12){\makebox(0,0){$v't_2$}}
\end{picture}
\end{center}
\caption{Cusp: the 2-leg webs.}
\label{F:cusp}
\end{figure}
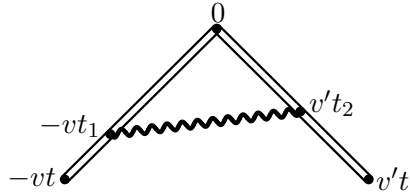

Now we consider the Green function ${<}T\{h_v^+(x)J(0)h_{v'}(x')\}{>}$.
Up to obvious $\delta$ functions similar to~(\ref{Intro:Sh}),
it is the broken Wilson line $W(\varphi)$
from $x=-vt$ to $0$ and then to $x'=v't'$.
Renormalization constants cannot depend on kinematics of Green functions
we choose to calculate, and so we choose $t' = t$ to have a single-scale problem.
We have
\begin{equation}
\log\frac{W(\varphi)}{W(0)} = \sum(w_i(\varphi)-w_i(0))\,,
\label{Gamma:exp}
\end{equation}
where the sum runs over all single-web diagrams.
Diagrams in which all photon vertices are to the left (or to the right) of the $J$ vertex
cancel in $w_i(\varphi)-w_i(0)$.
The remaining 2-leg webs are shown in fig.~\ref{F:cusp}.
At 4 loops 4-leg webs appear;
they have been calculated (at $\varphi\ll1$) in~\cite{Grozin:2017css}.

The the $L$-loop $n_l^1$ contribution is ($L\ge2$)
\begin{equation}
w_L(\varphi) = - \tilde{\Pi}_{L-1} n_l A_0^L e^{L\gamma\varepsilon}
\int_0^t dt_1 \int_0^t dt_2\,v_\mu v'_\nu \bar{D}_{L-1}^{\mu\nu}(vt_1+v't_2)\,,
\label{Gamma:wL}
\end{equation}
where $\bar{D}_L^{\mu\nu}(x)$ is given by~(\ref{gammah:Dx}).
We can write it, together with the 1-loop Landau-gauge contribution, in the form
\begin{equation}
w_1(\varphi) = V_1(\varphi) A\,,\quad
w_L(\varphi) = V_L(\varphi) \tilde{\Pi}_{L-1} n_l A^L\,,
\label{Gamma:w}
\end{equation}
where
\begin{equation}
V_L(\varphi) = 4 \frac{\Gamma(1-u)}{\Gamma(1+u-\varepsilon)} e^{-(2L-1)\gamma\varepsilon}
\left[ - I_1(\varphi) \cosh\varphi
+ \frac{u I_1(\varphi) \cosh\varphi - (1-u) I_2(\varphi)}{2(1+u-\varepsilon)}
\right]\,,
\label{Gamma:VL}
\end{equation}
$u=L\varepsilon$.
The integrals $I_{1,2}(\varphi)$ are
\begin{align}
I_1(\varphi) ={}& \int_0^1 dt_1 \int_0^1 dt_2 (e^{\varphi/2} t_1 + e^{-\varphi/2} t_2)^{-1+u} (e^{-\varphi/2} t_1 + e^{\varphi/2} t_2)^{-1+u}
\nonumber\\
{}={}& \frac{e^{-2 u \varphi}}{4 u^2 \sinh\varphi} \bigl(g_1(\varphi) - g_2(\varphi)\bigr)\,,
\label{Gamma:I1}\\
I_2(\varphi) ={}& \int_0^1 dt_1 \int_0^1 dt_2 (e^{\varphi/2} t_1 + e^{-\varphi/2} t_2)^u (e^{-\varphi/2} t_1 + e^{\varphi/2} t_2)^{-2+u}
\nonumber\\
{}={}& \frac{1}{2 u (1-u)} \left[1 + \frac{e^{-2 u \varphi}}{2 \sinh\varphi} \bigl(e^{-\varphi} g_1(\varphi) - e^\varphi g_2(\varphi)\bigr)\right]\,,
\label{Gamma:I2}\\
I_1(0) ={}& I_2(0) = \frac{2 - 2^{2u}}{2 u (1-2u)}\,,
\label{Gamma:I0}\\
g_1(\varphi) ={}& (e^\varphi+1)^{2u} f_1(1-e^\varphi) - f_1(1-e^{2\varphi})\,,\quad
g_2(\varphi) = (e^\varphi+1)^{2u} f_2(1-e^\varphi) - f_2(1-e^{2\varphi})\,,
\nonumber\\
f_1(x) ={}& \F{2}{1}{-2u,-u\\1-2u}{x} = 1 + 2 \Li2(x) u^2 + \mathcal{O}(u^3)\,,
\label{Gamma:f1}\\
f_2(x) ={}& \F{2}{1}{-2u,1-u\\1-2u}{x} = 1 + 2 \log(1-x) u + \left(\log^2(1-x) - 2 \Li2(x)\right) u^2 + \mathcal{O}(u^3)\,.
\label{Gamma:f2}
\end{align}

We obtain
\begin{align}
&V_L(\varphi) - V_L(0) = - 2 \frac{\varphi \coth\varphi - 1}{L \varepsilon} + \bar{V}(\varphi) + \mathcal{O}(\varepsilon)\,,
\label{Gamma:VLres}\\
&\bar{V}(\varphi) = \coth\varphi
\left[4 \Li2(1-e^{2\varphi}) - 4 \Li2(1-e^\varphi) + \varphi \left(4 \log(e^\varphi+1) + \varphi\right)\right]
\nonumber\\
&\quad{} + 2 \log(e^\varphi+1) - \varphi - 6 \log 2 + 4 = \bar{V}(-\varphi)\,.
\nonumber
\end{align}
Similarly to~(\ref{gammah:logW}), we substitute~(\ref{Gamma:VL}) into
\begin{equation*}
\log\frac{W(\varphi)}{W(0)} = \log Z_J + (\text{finite})
\end{equation*}
and re-express it via $\alpha(\mu_0)$ (it is sufficient to do this in the 1-loop term).
Note that $\bar{V}(\varphi)$ does not depend on $L$;
as a result, terms $\bar{\beta}_{L-1} \bar{V}(\varphi)$ cancel in $\log Z_J$
(in contrast to the first line of~(\ref{gammah:gamma}) where they contributed
because of the $1/L$ term in~(\ref{gammah:S})).
Differentiating $\log Z_J$ we obtain
\begin{align}
&\Gamma(\varphi) = 4 (\varphi \coth\varphi - 1) \frac{\alpha_s}{4\pi} \biggl[C_F
+ T_F n_l \sum_{L=1}^\infty \bar{\Pi}_L \left(C_F \frac{\alpha_s}{4\pi}\right)^L \biggr]
+ (\text{other color structures})
\nonumber\\
&{} = 4 (\varphi \coth\varphi - 1) C_F \frac{\alpha_s}{4\pi} \biggl\{1
+ T_F n_l \frac{\alpha_s}{4\pi} \biggl[
- \frac{20}{9}
+ \left(16 \zeta_3 - \frac{55}{3}\right) C_F \frac{\alpha_s}{4\pi}
\nonumber\displaybreak\\
&\quad{} - 2 \left(80 \zeta_5 - \frac{148}{3} \zeta_3 - \frac{143}{9}\right) \left(C_F \frac{\alpha_s}{4\pi}\right)^2
\nonumber\\
&\quad{} + \left(2240 \zeta_7 - 1960 \zeta_5 - 104 \zeta_3 + \frac{31}{3}\right) \left(C_F \frac{\alpha_s}{4\pi}\right)^3
+ \mathcal{O}(\alpha_s^4) \biggr] \biggr\}
\nonumber\\
&\quad{} + (\text{other color structures})\,.
\label{Gamma:Gamma}
\end{align}
Thus we have reproduced the 3-loop $C_F^2 T_F n_l$ term in~\cite{Grozin:2014hna,Grozin:2015kna}.
The coefficient of $2 T_F n_l C_F^3 (\alpha_s/(4\pi))^4$ in the light-like cusp anomalous dimension $\Gamma_l$ is
\begin{equation*}
- 4 \left(80 \zeta_5 - \frac{148}{3} \zeta_3 - \frac{143}{9}\right) \approx - 31.055431\,,
\end{equation*}
in perfect agreement with the numerical result $-31.00\pm0.4$ (Table~2 in~\cite{Moch:2017uml}).

The $C_F^{L-1} T_F n_l$ terms in the quark--antiquark potential in Coulomb gauge are given by a single Coulomb-gluon propagator:
\begin{align}
&V(\vec{q}^{\,}) = - \frac{4 \pi \alpha_s}{\vec{q}^{\,2}} \biggl[ C_F + T_F n_l \sum_{L=1}^\infty \bar{\Pi}_L \left(C_F \frac{\alpha_s}{4\pi}\right)^L \biggr]
+ (\text{other color structures})
\nonumber\\
&{} = - \frac{4 \pi \alpha_s}{\vec{q}^2} \biggl\{ C_F + T_F n_l \frac{\alpha_s}{4\pi} \biggl[ - \frac{20}{9}
+ \biggl(16 \zeta_3 - \frac{55}{3}\biggr) \frac{\alpha_s}{4\pi}
\nonumber\\
&\quad{} - 2 \biggl(80 \zeta_5 - \frac{148}{3} \zeta_3 - \frac{143}{9}\biggr) \biggl(C_F \frac{\alpha_s}{4\pi}\biggr)^2
\nonumber\\
&\quad{} + \biggl(2240 \zeta_7 - 1960 \zeta_5 - 104 \zeta_3 + \frac{31}{3} \biggr) \biggl(C_F \frac{\alpha_s}{4\pi}\biggr)^3
+ \mathcal{O}(\alpha_s^4)  \biggr] \biggr\}
\nonumber\\
&{} + (\text{other color structures})\,,
\label{Gamma:V}
\end{align}
where $\alpha_s$ is taken at $\mu^2 = \vec{q}^{\,2}$.
The terms up to $\alpha_s^4$ agree with~\cite{Smirnov:2008pn}.
The cusp anomalous dimension at Euclidean angle $\varphi_E = \pi - \delta$, $\delta\to0$,
is related to the quark--antiquark potential~\cite{Kilian:1993nk}
\begin{equation}
\delta\,\Gamma(\pi-\delta)\Big|_{\delta\to0} = \frac{\vec{q}^{\,2} V(\vec{q}^{\,})}{4\pi}\,;
\label{Gamma:delta}
\end{equation}
this relation follows from conformal invariance,
and in QCD it is broken by extra terms proportional to coefficients of the $\beta$ function~\cite{Grozin:2014hna,Grozin:2015kna}.
Comparing~(\ref{Gamma:Gamma}) with~(\ref{Gamma:V}), we see that the relation~(\ref{Gamma:delta})
for the $C_F^{L-1} T_F n_l$ color structures is valid to all orders in $\alpha_s$.

\section{QED results}
\label{S:QED}

The 4-loop anomalous dimension of the QED Bloch--Nordsieck field is now known
completely analytically.
Adding terms with higher powers of $n_l$ from~\cite{Grozin:2015kna,Grozin:2016ydd}
and the 4-loop contribution of the webs with 4 legs~\cite{Grozin:2017css}, we obtain
\begin{align}
\gamma_h ={}& 2 (a-3) \frac{\alpha}{4\pi}
+ \frac{32}{3} n_l \left(\frac{\alpha}{4\pi}\right)^2
+ \left[ - 6 \left(16 \zeta_3 - 17\right) + \frac{160}{27} n_l \right]
n_l \left(\frac{\alpha}{4\pi}\right)^3
\nonumber\\
&{} + \biggl[ 16 \left(40 \zeta_5 + \frac{32}{3} \pi^2 \zeta_3 - 21 \zeta_3 - \frac{32}{3} \pi^2 - \frac{35}{3}\right)
\nonumber\\
&\hphantom{{}+\biggl[\biggr.}
- 32 \left(\frac{\pi^4}{15} - 12 \zeta_3 + \frac{103}{27}\right) n_l
- \frac{256}{9} \left(\zeta_3 - \frac{1}{3}\right) n_l^2
\biggr] n_l \left(\frac{\alpha}{4\pi}\right)^4\,.
\label{QED:gammah}
\end{align}

Adding terms with higher powers of $n_l$~\cite{Grozin:2015kna,Grozin:2016ydd}
and the 4-legs webs contribution~\cite{Grozin:2017css,Bruser:2018aud} (known only up to $\varphi^6$)
to~(\ref{Gamma:Gamma}),
we obtain the QED cusp anomalous dimension up to 4 loops
\begin{align}
\Gamma(\varphi) ={}& 4 (\varphi \coth\varphi - 1) \frac{\alpha}{4\pi} \biggl\{ 1 + n_l \frac{\alpha}{4\pi}
\biggl[ - \frac{20}{9}
+ \left(16 \zeta_3 - \frac{55}{3} - \frac{16}{27} n_l\right) \frac{\alpha}{4\pi}
\nonumber\\
&\quad{} - \frac{8}{9} \biggl(\frac{2}{5} \pi^4 - 80 \zeta_3 + \frac{299}{9}
- \frac{8}{3} \biggl(2 \zeta_3 - \frac{1}{3}\biggr) n_l\biggr) n_l \left(\frac{\alpha}{4\pi}\right)^2 \biggr] \biggr\}
\nonumber\\
&{} - \frac{8}{3} \varphi^2 \biggl[80 \zeta_5 + \frac{128}{3} \pi^2 \zeta_3 - \frac{40}{9} \pi^4
- \frac{148}{3} \zeta_3 - \frac{80}{9} \pi^2 - \frac{143}{9}
\nonumber\\
&\quad{} + \frac{1}{3} \biggl(112 \zeta_5 + \frac{512}{75} \pi^2 \zeta_3 - \frac{392}{225} \pi^4
- \frac{6076}{75} \zeta_3 + \frac{1256}{225} \pi^2 + \frac{2371}{225}\biggr) \varphi^2
\nonumber\\
&\quad{} - \frac{2}{3} \biggl( \frac{304}{49} \zeta_5 + \frac{512}{1225} \pi^2 \zeta_3 - \frac{42004}{11025} \zeta_3
- \frac{3368}{33075} \pi^4 + \frac{10664}{33075} \pi^2 - \frac{9341}{33075} \biggr) \varphi^4
\nonumber\\
&\quad{} + \mathcal{O}(\varphi^6) \biggr] n_l \left(\frac{\alpha}{4\pi}\right)^4\,.
\label{QED:Gamma}
\end{align}
The $n_l \alpha^4$ term is known only up to $\varphi^6$;
\begin{equation*}
\varphi \coth\varphi - 1 = \frac{\varphi^2}{3} \left(1 - \frac{\varphi^2}{15} + \frac{2}{315} \varphi^4 + \mathcal{O}(\varphi^6)\right)\,.
\end{equation*}

\acknowledgments

I am grateful to J.\,M.~Henn and M.~Stahlhofen for hospitality in Mainz and useful discussions;
to M.~Steinhauser and P.~Marquard for discussing~\cite{Marquard:2018rwx};
to A.~Vogt for discussing~\cite{Moch:2017uml}.
The work has been supported by the PRISMA cluster of excellence, JGU Mainz,
and partially by the Russian Ministry of Education and Science.


\begin{thebibliography}{99}

\bibitem{Marquard:2018rwx}
  P.~Marquard, A.\,V.~Smirnov, V.\,A.~Smirnov and M.~Steinhauser,
  \textit{Four-loop wave function renormalization in QCD and QED},
  Phys.\ Rev.\ \textbf{D 97} (2018) no.~5, 054032
  [arXiv:1801.08292 [hep-ph]].

\bibitem{Moch:2017uml}
  S.~Moch, B.~Ruijl, T.~Ueda, J.\,A.\,M.~Vermaseren and A.~Vogt,
  \textit{Four-loop non-singlet splitting functions in the planar limit and beyond},
  JHEP \textbf{1710} (2017) 041
  [arXiv:1707.08315 [hep-ph]].

\bibitem{Grozin:2017css}
  A.~Grozin, J.~Henn and M.~Stahlhofen,
  \textit{On the Casimir scaling violation in the cusp anomalous dimension at small angle},
  JHEP \textbf{1710} (2017) 052
  [arXiv:1708.01221 [hep-ph]].

\bibitem{Bruser:2018aud}
  R.~Br\"{u}ser, A.~G.~Grozin, J.~M.~Henn and M.~Stahlhofen,
  arXiv:1807.05145 [hep-ph].

\bibitem{Manohar:2000dt}
  A.\,V.~Manohar and M.\,B.~Wise,
  \textit{Heavy quark physics},
  Camb.\ Monogr.\ Part.\ Phys.\ Nucl.\ Phys.\ Cosmol.\ \textbf{10} (2000) 1.

\bibitem{Grozin:2004yc}
  A.\,G.~Grozin,
  \textit{Heavy quark effective theory},
  Springer Tracts Mod.\ Phys.\ \textbf{201} (2004) 1.

\bibitem{Grozin:2013bva}
  A.\,G.~Grozin,
  \textit{Introduction to effective field theories. 3. Bloch--Nordsieck effective theory, HQET},
  arXiv:1305.4245 [hep-ph].

\bibitem{Grozin:2010wa}
  A.\,G.~Grozin,
  \textit{Matching heavy-quark fields in QCD and HQET at three loops},
  Phys.\ Lett.\ \textbf{B 692} (2010) 161
  [arXiv:1004.2662 [hep-ph]].

\bibitem{Melnikov:2000zc}
  K.~Melnikov and T.~van Ritbergen,
  \textit{The three-loop on-shell renormalization of QCD and QED},
  Nucl.\ Phys.\ \textbf{B 591} (2000) 515
  [hep-ph/0005131].

\bibitem{Chetyrkin:2003vi}
  K.\,G.~Chetyrkin and A.\,G.~Grozin,
  \textit{Three-loop anomalous dimension of the heavy--light quark current in HQET},
  Nucl.\ Phys.\ \textbf{B 666} (2003) 289
  [hep-ph/0303113].

\bibitem{Broadhurst:1994se}
  D.\,J.~Broadhurst and A.\,G.~Grozin,
  \textit{Matching QCD and HQET heavy--light currents at two loops and beyond},
  Phys.\ Rev.\ \textbf{D 52} (1995) 4082
  [hep-ph/9410240].

\bibitem{Grozin:2015kna}
  A.~Grozin, J.\,M.~Henn, G.\,P.~Korchemsky and P.~Marquard,
  \textit{The three-loop cusp anomalous dimension in QCD and its supersymmetric extensions},
  JHEP \textbf{1601} (2016) 140
  [arXiv:1510.07803 [hep-ph]].

\bibitem{Grozin:2016ydd}
  A.~Grozin,
  \textit{Leading and next-to-leading large-$n_f$ terms in the cusp anomalous dimension and quark--antiquark potential},
  PoS LL \textbf{2016} (2016) 053
  [arXiv:1605.03886 [hep-ph]].

\bibitem{Lee:2013sx}
  R.~Lee, P.~Marquard, A.\,V.~Smirnov, V.\,A.~Smirnov and M.~Steinhauser,
  \textit{Four-loop corrections with two closed fermion loops to fermion self energies and the lepton anomalous magnetic moment},
  JHEP \textbf{1303} (2013) 162
  [arXiv:1301.6481 [hep-ph]].

\bibitem{Grozin:2014hna}
  A.~Grozin, J.\,M.~Henn, G.\,P.~Korchemsky and P.~Marquard,
  \textit{Three-loop cusp anomalous dimension in QCD},
  Phys.\ Rev.\ Lett.\ \textbf{114} (2015) no.~6, 062006
  [arXiv:1409.0023 [hep-ph]].

\bibitem{Korchemsky:1987wg}
  G.\,P.~Korchemsky and A.\,V.~Radyushkin,
  \textit{Renormalization of the Wilson loops beyond the leading order},
  Nucl.\ Phys.\ \textbf{B 283} (1987) 342.

\bibitem{Beneke:1995pq}
  M.~Beneke and V.\,M.~Braun,
  \textit{Power corrections and renormalons in Drell--Yan production},
  Nucl.\ Phys.\ \textbf{B 454} (1995) 253
  [hep-ph/9506452].

\bibitem{Bagan:1993js}
  E.~Bagan and P.~Gosdzinsky,
  \textit{Two-loop renormalization scale dependence of the Isgur--Wise function},
  Phys.\ Lett.\ \textbf{B 305} (1993) 157.

\bibitem{Henn:2016men}
  J.\,M.~Henn, A.\,V.~Smirnov, V.\,A.~Smirnov and M.~Steinhauser,
  \textit{A planar four-loop form factor and cusp anomalous dimension in QCD},
  JHEP \textbf{1605} (2016) 066
  [arXiv:1604.03126 [hep-ph]].

\bibitem{Davies:2016jie}
  J.~Davies, A.~Vogt, B.~Ruijl, T.~Ueda and J.\,A.\,M.~Vermaseren,
  \textit{Large-$n_f$ contributions to the four-loop splitting functions in QCD},
  Nucl.\ Phys.\ \textbf{B 915} (2017) 335
  [arXiv:1610.07477 [hep-ph]].

\bibitem{Lee:2016ixa}
  J.~Henn, R.\,N.~Lee, A.\,V.~Smirnov, V.\,A.~Smirnov and M.~Steinhauser,
  \textit{Four-loop photon quark form factor and cusp anomalous dimension in the large-$N_c$ limit of QCD},
  JHEP \textbf{1703} (2017) 139
  [arXiv:1612.04389 [hep-ph]].

\bibitem{Gatheral:1983cz}
  J.\,G.\,M.~Gatheral,
  \textit{Exponentiation of eikonal cross-sections in nonabelian gauge theories},
  Phys.\ Lett.\ \textbf{B 133} (1983) 90.

\bibitem{Frenkel:1984pz}
  J.~Frenkel and J.\,C.~Taylor,
  \textit{Nonabelian eikonal exponentiation},
  Nucl.\ Phys.\ \textbf{B 246} (1984) 231.

\bibitem{Gorishnii:1990kd}
  S.\,G.~Gorishny, A.\,L.~Kataev, S.\,A.~Larin and L.\,R.~Surguladze,
  \textit{The analytical four-loop corrections to the QED $\beta$-function in the MS scheme and to the QED $\psi$-function: Total reevaluation},
  Phys.\ Lett.\ \textbf{B 256} (1991) 81.

\bibitem{Ruijl:2017eht}
  B.~Ruijl, T.~Ueda, J.\,A.\,M.~Vermaseren and A.~Vogt,
  \textit{Four-loop QCD propagators and vertices with one vanishing external momentum},
  JHEP \textbf{1706} (2017) 040
  [arXiv:1703.08532 [hep-ph]].

\bibitem{Smirnov:2008pn}
  A.\,V.~Smirnov, V.\,A.~Smirnov and M.~Steinhauser,
  \textit{Fermionic contributions to the three-loop static potential},
  Phys.\ Lett.\ \textbf{B 668} (2008) 293
  [arXiv:0809.1927 [hep-ph]].

\bibitem{Kilian:1993nk}
  W.~Kilian, T.~Mannel and T.~Ohl,
  \textit{Unimagined imaginary parts in heavy quark effective field theory},
  Phys.\ Lett.\ \textbf{B 304} (1993) 311
  [hep-ph/9303224].

\end{thebibliography}
\end{document}